\documentclass[preprint,review,10pt]{elsarticle}
 
\usepackage[top=4.3cm,right=4.8cm,bottom=4.3cm,left=4.8cm]{geometry}
\usepackage{setspace}
\usepackage{amssymb}
\usepackage{amsmath}
 
\usepackage{subcaption}
\usepackage{url}
\usepackage{booktabs}
\usepackage{adjustbox}
\usepackage{tabularx}
\usepackage{makecell}

\usepackage{xurl} 

\usepackage{multirow}
\usepackage{float}
\usepackage[table]{xcolor}
\usepackage{colortbl}

\journal{Pattern Recognition}

\begin{document}

\begin{frontmatter}

\title{Geo-ATBench: A Benchmark for Geospatial Audio Tagging with Geospatial Semantic Context}

\author[ox]{Yuanbo Hou\fnref{eq}\corref{cor1}}
\author[xjtlu]{Yanru Wu\fnref{eq}}
\author[kth]{Qiaoqiao Ren}
\author[xjtlu]{Shengchen Li}
\author[ox]{Stephen Roberts}
\author[ugent]{Dick Botteldooren}
 
 
\affiliation[ox]{organization={Machine Learning Research Group, Engineering Science, University of Oxford}, country={UK}}
\affiliation[xjtlu]{organization={Department of Intelligent Science, Xi'an Jiaotong-Liverpool University}, country={China}}
\affiliation[kth]{organization={EECS, KTH Royal Institute of Technology}, country={Sweden}}
\affiliation[ugent]{organization={WAVES Research Group, Information Technology, Ghent University}, country={Belgium}}

\fntext[eq]{Equal contribution.}
\cortext[cor1]{Corresponding author: Yuanbo Hou, Machine Learning Research Group, University of Oxford, UK. Email: Yuanbo.Hou@eng.ox.ac.uk}

\begin{abstract} 
Environmental sound understanding in computational auditory scene analysis (CASA) is often formulated as an audio-only recognition problem. This formulation leaves a persistent drawback in multi-label audio tagging (AT): acoustic similarity can make certain events difficult to separate from waveforms alone. In such cases, disambiguating cues often lie outside the waveform. Geospatial semantic context (GSC), derived from geographic information system data, e.g., points of interest (POI), provides location-tied environmental priors that can help reduce this ambiguity.
A systematic study of this direction is enabled through the proposed geospatial audio tagging (Geo-AT) task, which conditions multi-label sound event tagging on GSC alongside audio. To benchmark Geo-AT, the Geo-ATBench dataset is introduced as a polyphonic audio benchmark with geographical annotations, containing 10.71 hours of real-world audio across 28 event categories; each clip is paired with a POI-derived GSC representation constructed from 11 semantic context categories. Furthermore, GeoFusion-AT is proposed as a unified geo-audio fusion framework that evaluates feature-level, representation-level, and decision-level fusion on three representative audio backbones, with audio-only and GSC-only baselines.  
Experiments show that incorporating GSC generally improves AT performance, especially on acoustically confounded labels, indicating that geospatial semantics can provide an effective prior beyond audio alone. 
A crowdsourced listening study with 10 participants on 579 samples shows that there is no significant difference in performance between the models on the Geo-ATBench labels and on aggregated human labels, supporting Geo-ATBench as a human-aligned benchmark. Overall, the proposed Geo-AT task, the open benchmark Geo-ATBench, and the reproducible geo-audio fusion framework GeoFusion-AT provide a solid foundation for studying audio tagging with geospatial semantic context within the CASA community. For the dataset, source code, and models,
please see the project homepage (\textcolor{blue}{\url{https://github.com/WuYanru2002/Geo-ATBench}}).
\end{abstract}
  
\begin{keyword}

Computational auditory scene analysis \sep Multi-label audio tagging \sep Geospatial semantic context \sep Points of interest \sep Multimodal fusion
\end{keyword}

\end{frontmatter}
 
\section{Introduction}
\label{section_Introduction}

Environmental sound understanding is one of the core goals of computational auditory scene analysis (CASA)~\cite{ virtanen2018computational}. In many practical applications, the target output is multi-label audio tagging (AT) \cite{DCASE2016}, where each recording may contain multiple sound events and the system predicts the set of event labels. AT supports applications such as acoustic surveillance~\cite{lopatka2015detection}, smart-city sensing~\cite{zhang2023automatic}, multimedia retrieval~\cite{wold1996content}, and intelligent domestic assistants~\cite{vacher2020audio}.

Despite strong progress in deep learning models for environmental audio, AT is commonly treated as an audio-only recognition problem~\cite{Mesaros2021tutorial, Purwins2019deeplearning}. Recent AT backbones, including convolutional neural networks (CNNs) and Transformers, learn powerful acoustic representations from time-frequency features such as Mel spectrograms~\cite{Gong2021ast}. However, a persistent drawback remains. 
Acoustic similarity can make certain events difficult to distinguish from waveforms alone, especially when different sources produce highly similar time-frequency patterns~\cite{Salamon2017deep, morfi2018deep}.  
In such cases, disambiguating cues often lie beyond the waveform~\cite{Heittola2013context, primus2024fusing}. 
A key source of such cues is the physical environment in which sound occurs. Sound events are produced by sources embedded in specific places, and their occurrence is shaped by location-tied environmental factors~\cite{Bregman1994auditory}.  
Location-tied conditions can induce systematic associations between event labels and geospatial semantic context (GSC)~\cite{diez2023noisensedb}. GSC can therefore provide complementary cues when waveforms alone are ambiguous.

This work focuses on sound source-associated GSC, which refers to location-tied environmental priors derived from geographic information systems data, such as points of interest (POI) \cite{yuan2013discovering}. 
Compared with raw GPS coordinates, POI-derived GSC provides structured semantic descriptions of the physical environment surrounding sound sources that can be aligned with audio representations~\cite{dalei2025geographic}.  
Progress in this direction remains limited by the lack of standardized tasks and benchmark datasets that pair audio with reliable, structured GSC under reproducible evaluation~\cite{cartwright2020sonyc}. 
Recent mobile recording devices and location-aware media platforms increasingly associate recordings with geographic coordinates~\cite{cartwright2020sonyc}, making relevant audio-GSC pairs increasingly accessible. This trend creates a timely opportunity to investigate how to leverage GSC to support multi-label AT in the real world.

To address the gap that AT is often formulated without sound source-associated location-tied GSC, this paper proposes the geospatial audio tagging (Geo-AT) task, which conditions multi-label AT on GSC alongside audio recordings. Geo-AT aims to assess whether location-tied environmental priors help disambiguate events that are difficult to distinguish from audio alone. 
To benchmark Geo-AT, we release the Geo-ATBench dataset, a geographically annotated polyphonic audio benchmark containing 3,854 clips with 28 event labels; each clip is paired with a GSC representation constructed from POI semantics over 11 context categories, enabling reproducible studies of how geospatial semantics interact with acoustic representations in multi-label AT.

The proposed benchmark design of Geo-ATBench does not specify how GSC should be integrated into AT models \cite{DCASE2016, lopatka2015detection, Kumar2016audio}, and different integration choices may lead to different outcomes. Therefore, GeoFusion-AT is introduced as a unified geo-audio fusion framework for the proposed Geo-AT task to benchmark representative fusion strategies and to report reference results on Geo-ATBench. Specifically, GeoFusion-AT evaluates three typical fusion strategies, feature-level, representation-level, and decision-level fusion, across three representative audio backbones, the CNN-based pretrained audio neural networks (PANNs) \cite{Kong2020panns}, the Transformer-based audio spectrogram Transformer (AST) \cite{Gong2021ast}, and contrastive language-audio pretraining (CLAP) \cite{elizalde2023clap}. Audio-only and GSC-only baselines are included to isolate the contribution of each modality and to identify when fusion improves performance beyond either input alone.

The main contributions are: 1) Geo-AT is introduced as a standardized task formulation for multi-label audio tagging in CASA that integrates audio with geospatial semantic context (GSC); 
2) Geo-ATBench is released as an open benchmark for reproducible Geo-AT evaluation, containing 3,854 real-world polyphonic audio clips annotated with 28 event labels, where each clip is paired with a GSC representation constructed from POI semantics over 11 semantic context categories;
3) GeoFusion-AT is introduced as a unified geo-audio fusion framework that benchmarks representative fusion strategies across representative audio backbones on Geo-ATBench to report reference results; 4) A crowdsourced listening study with 10 participants on 579 samples is conducted, showing that model performance is comparable when evaluated against Geo-ATBench labels and aggregated human labels, supporting Geo-ATBench as a human-aligned benchmark.
We have released the dataset, code, and models.

The rest of this paper is organized as follows. Section 2 reviews related work to position Geo-AT. Section 3 formalizes the Geo-AT task. Section 4 describes the Geo-ATBench dataset. Section 5 presents the GeoFusion-AT framework with fusion strategies based on representative audio backbones. Section 6 reports experimental results and analysis. Section 7 details the human evaluation study. Section 8 concludes the paper.

\section{Related Work}
\label{section_Related}
This section positions the proposed geospatial audio tagging (Geo-AT) task within prior work on multi-label audio tagging (AT), context-aware sound understanding, and POI-derived geospatial semantic context from geographic information systems. The discussion motivates the need for a standardized Geo-AT task under reproducible evaluation.

\subsection{Multi-Label Audio Tagging and Acoustic Ambiguity}

Multi-label AT is a central task in CASA~\cite{virtanen2018computational}, where an audio clip may contain polyphonic sound events, and the goal is to predict the set of event labels~\cite{DCASE2016}. Large-scale benchmarks and challenges~\cite{fonseca2021fsd50k, piczak2015esc} have driven steady progress in model architectures and backbones, such as CNN-based PANNs~\cite{Kong2020panns} and MobileNet~\cite{sandler2018mobilenetv2}, Transformer-based Hierarchical Token-Semantic Audio Transformer~\cite{chen2022hts} and AST~\cite{Gong2021ast}, with contrastive learning-based CLAP that aligns audio and language representations~\cite{elizalde2023clap}. These backbones have become common reference points for representation learning in AT tasks.
 
Despite architectural advances, AT in real-world conditions continues to face persistent ambiguity~\cite{Gemmeke2017audioset}. Polyphonic recordings often contain overlapping sources, and different events can produce similar time-frequency patterns~\cite{Salamon2017deep, morfi2018deep}, leading audio-only AT to struggle with confusable events and misclassification. 
External priors like sound source-associated GSC provide complementary cues by encoding location-tied environmental priors into a structured POI-derived semantic representation \cite{yuan2013discovering}, such as nearby place categories and their composition around the sound source.  
Location-tied GSC constrains the set of plausible events for a scene and can support disambiguation when acoustic evidence alone is insufficient.

\subsection{Context and Auxiliary Information for Sound Understanding}
 
Context-aware sound understanding extends AT by incorporating information beyond acoustic representations. Prior work~\cite{chen2020vggsound, kim2019audiocaps} can be divided into two groups, distinguished by whether the additional signal is time-aligned with the audio. 
One group~\cite{zhou2018visual} uses paired sensory streams, where video frames or other time-aligned inputs are available together with the audio. Another group~\cite{yang2022avqa} uses auxiliary metadata that is linked to the recording environment but is not time-aligned with the audio signal.

Geo-AT concerns the second group. Location-tied descriptors operate as scene-level priors and remain available in many deployments. Existing studies~\cite{shaikh2024multimodal} that incorporate auxiliary metadata vary in metadata representation, audio-metadata pairing rules, data splits, and reporting practice. Audio-only and metadata-only baselines are not always reported. These inconsistencies limit reproducible comparison across studies and motivate a standardized task for evaluating auxiliary metadata in multi-label AT.
 
\subsection{POI-Derived Geospatial Semantic Context (GSC)}

Geospatial information has become increasingly available in audio collections due to mobile recording devices and location-aware media platforms that associate recordings with geographic coordinates~\cite{cartwright2020sonyc}. Several datasets include geographic or location-related annotations, enabling spatial analyses of urban sound environments and regional differences~\cite{diez2023noisensedb}. However, geospatial information is usually used for organization, mapping, or descriptive analysis rather than as an explicit model input for sound event recognition~\cite{quinn2022soundscape}.
  
Points of interest (POI) in geographic information systems translate location into interpretable semantic descriptors. POI encodes nearby places and compositions, representing location-tied environmental priors~\cite{yuan2013discovering}. POI-derived GSC contains scene-level descriptors that can be paired with audio recordings. However, prior work rarely formalizes POI-derived GSC as part of the AT task. The lack of consistent task definitions and benchmarks makes it difficult to assess whether and how geospatial semantics should be integrated.

\subsection{Positioning Geo-AT}

Taken together, prior work leaves AT largely audio-only and rarely evaluates POI-derived GSC as a task input under reproducible protocols. The missing piece is a standardized Geo-AT task definition and a benchmark that enables controlled comparisons. The Geo-AT task addresses this gap by defining AT conditioned on sound source-associated, POI-derived GSC alongside audio, enabling controlled evaluation of geospatial priors in AT tasks.

\section{The proposed Geospatial Audio Tagging (Geo-AT) task}
\label{section_task}
  
Geospatial audio tagging (Geo-AT) formalizes AT conditioned on sound source-associated geospatial semantic context (GSC) derived from geographic information systems resources, such as Points of Interest (POI). Geo-AT is a multimodal learning task that enables controlled study of how POI-derived GSC interact with acoustic representations in AT tasks \cite{DCASE2016, Kumar2016audio}.

\subsection{Problem Definition}

Given each recording is represented by an acoustic representation $\mathbf{A}$ and a GSC vector $\mathbf{g} \in \mathbb{R}^{D_{\text{GSC}}}$ constructed from  geographic information systems,
Geo-AT uses a paired input $(\mathbf{A}, \mathbf{g})$. Geo-AT assumes that $\mathbf{g}$ is available as recording metadata at inference time, alongside $\mathbf{A}$. The learning objective is to predict the set of event labels present in the clip.
Let $\mathcal{Y}$ denote the event label set. The target for each clip is a multi-label vector $\mathbf{y} \in \{0,1\}^{|\mathcal{Y}|}$, where $\mathbf{y}_k = 1$ indicates the presence of event $k$ in the clip. Geo-AT aims to learn a function
$f: (\mathbf{A}, \mathbf{g}) \to \mathbf{y}$, 
where $\mathbf{g}$ encodes information about the surrounding environment through POI-derived semantic descriptors (e.g., proximity to beaches, highways, train stations, residential areas, or industrial facilities). Geo-AT does not prescribe a specific integration mechanism between $\mathbf{A}$ and $\mathbf{g}$, leaving model design choices open for evaluation under a shared task definition.
 
\subsection{Task scope}
 
Geo-AT focuses on multi-label tagging rather than single-label classification, emphasizing label prediction under polyphonic conditions, where multiple events may co-occur in a clip. The purpose of Geo-AT is not to replace the AT task, but to study when and how spatial evidence complements audio representations, particularly for acoustically confusable events and polyphonic overlap. Geo-AT is motivated by the use of contextual knowledge in auditory perception and location-tied metadata in real deployments~\cite{boes2018machine}. 
Geo-AT provides a framework for building and evaluating more robust machine listening systems in geographically diverse environments, including urban noise monitoring, context-aware assistive hearing, and scalable acoustic surveillance~\cite{cartwright2020sonyc}~\cite{wang2015context}.

\section{The benchmark dataset for the Geo-AT task: Geo-ATBench}\label{section_dataset}

\subsection{Data collection}

The audio recordings for Geo-ATBench are sourced from Freesound.org \cite{fonseca2017freesound}, a public repository of user-contributed sounds, as well as from the dataset presented in \cite{boes2018machine}, which includes audio files with GPS information and a diverse range of sound events. Audio clips were selected based on the inclusion of geotagging information, specifically latitude and longitude coordinates provided by the uploaders, and underwent careful manual review of coordinate validity and obvious mismatches between tags and location for quality control.

\subsection{Sound event and GSC annotation}\label{coarse_grained_3class}

\begin{figure}[t]
\centering
\setlength{\abovecaptionskip}{0.05cm}    
\setlength{\belowcaptionskip}{-0.3cm} 
\includegraphics[width=0.6\columnwidth]{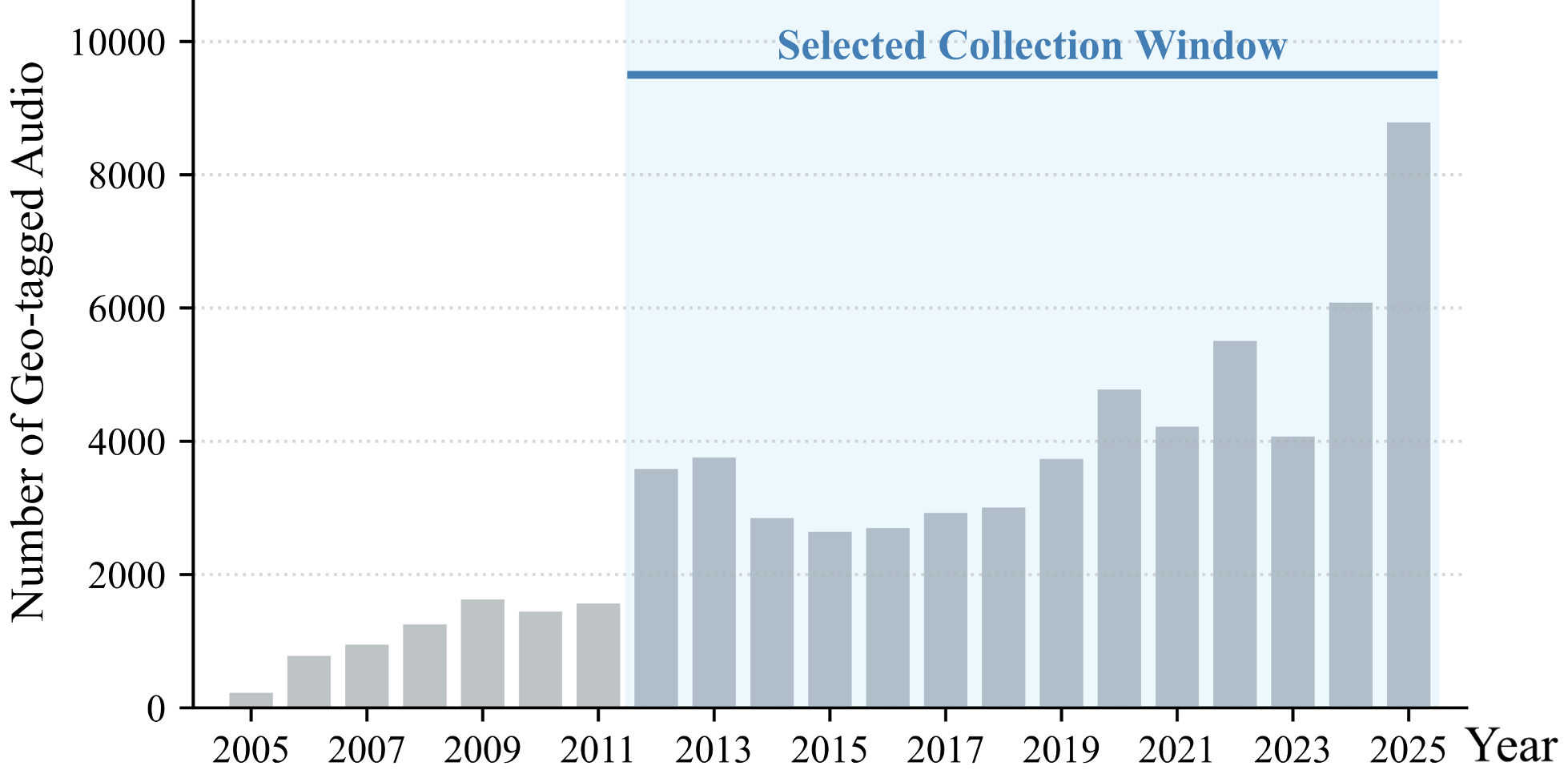}
\caption{The number of recordings with GPS information uploaded to Freesound each year.}
\label{fig:num_per_year}
\end{figure} 

\textbf{GSC construction:} For recordings sourced from Freesound, we specifically select data spanning from 2012 to 2025. This temporal filtering is applied because the scale of geo-tagged audio prior to 2012 is relatively limited, as shown in Fig.~\ref{fig:num_per_year}, and the geographical information of regions may differ across long time spans. The GPS coordinates of each recording were obtained from Freesound or the original dataset for others. These coordinates are used to query the OpenStreetMap (OSM) geospatial database via the Overpass API~\cite{OSM}. For each recording with GPS coordinates, a square with a fixed side length is drawn around the location, and OSM entities within this square are identified based on 11 OSM feature keys, covering categories such as land use, amenities, and natural. While a circular region may be conceptually aligned with the isotropic nature of sound propagation, a square region is adopted to enable efficient bounding-box queries within standard OSM-based geographic information systems. This choice provides a computationally practical approximation of the local acoustic environment while maintaining spatial consistency across samples. The resulting GSC representation is a POI-derived semantic descriptor extracted from these OSM annotations and used as the location-tied input described in Section~\ref{section_task}. The square side length and the 11 feature keys are the same for all clips to keep GSC extraction consistent across the dataset.

\begin{table}[t]
\centering
\footnotesize
\setlength{\tabcolsep}{2pt} 
\setlength{\abovecaptionskip}{0.1cm}   
	\setlength{\belowcaptionskip}{-0.4cm}  
\renewcommand{\arraystretch}{1} 

\definecolor{naturegreen}{HTML}{E8F5E9}
\definecolor{humanorange}{HTML}{FFF3E0}
\definecolor{thingblue}{HTML}{E3F2FD}
\definecolor{headergray}{HTML}{F5F5F5}

\resizebox{0.85\textwidth}{!}{
\begin{tabularx}{\textwidth}{@{} l r r| l r r| l r r @{}}
\toprule
\rowcolor{headergray}
\textbf{Class} & \textbf{Cnt.} & \textbf{Dur.(s)} &
\textbf{Class} & \textbf{Cnt.} & \textbf{Dur.(s)} &
\textbf{Class} & \textbf{Cnt.} & \textbf{Dur.(s)} \\
\midrule

\multicolumn{3}{c|}{\cellcolor{naturegreen}\textbf{Natural Sounds}} &
\multicolumn{3}{c|}{\cellcolor{humanorange}\textbf{Human Sounds}} &
\multicolumn{3}{c}{\cellcolor{thingblue}\textbf{Sounds of Things}} \\
\midrule

\cellcolor{naturegreen}Bird sounds     & 1024 & 8191 &
\cellcolor{humanorange}Speech          & 794  & 5133 &
\cellcolor{thingblue}Car               & 463  & 3068 \\

\cellcolor{naturegreen}Crickets        & 343  & 3091 &
\cellcolor{humanorange}Footsteps       & 288  & 2225 &
\cellcolor{thingblue}Plane             & 340  & 3092 \\

\cellcolor{naturegreen}Falling water   & 325  & 2922 &
\cellcolor{humanorange}Music Instru.   & 188  & 1593 &
\cellcolor{thingblue}Train             & 165  & 1291 \\

\cellcolor{naturegreen}Flowing water   & 319  & 2774 &
\cellcolor{humanorange}Music           & 144  & 1330 &
\cellcolor{thingblue}Bell              & 121  & 835  \\

\cellcolor{naturegreen}Waves           & 307  & 2754 &
\cellcolor{humanorange}Singing         & 81   & 624  &
\cellcolor{thingblue}Boat              & 115  & 927  \\

\cellcolor{naturegreen}Insects(Flying) & 137  & 824  &
\cellcolor{humanorange}Shout/Scream    & 79   & 249  &
\cellcolor{thingblue}Tram              & 111  & 731  \\

\cellcolor{naturegreen}Wind            & 83   & 737  &
\cellcolor{humanorange}Laughter        & 53   & 125  &
\cellcolor{thingblue}Vehicle horn      & 107  & 293  \\

                 &      &      &
                 &      &      &
\cellcolor{thingblue}Explosion         & 93   & 431  \\

                 &      &      &
                 &      &      &
\cellcolor{thingblue}Bus               & 74   & 461  \\

                 &      &      &
                 &      &      &
\cellcolor{thingblue}Siren             & 69   & 509  \\

                 &      &      &
                 &      &      &
\cellcolor{thingblue}Metro             & 63   & 454  \\

                 &      &      &
                 &      &      &
\cellcolor{thingblue}Helicopter        & 58   & 496  \\

                 &      &      &
                 &      &      &
\cellcolor{thingblue}Dog               & 209  & 969  \\

                 &      &      &
                 &      &      &
\cellcolor{thingblue}Truck             & 42   & 237  \\

\bottomrule
\end{tabularx}
}
\caption{Sound classes in Geo-ATBench, grouped by Natural, Human, and Thing. Dur. denotes the total duration (in seconds) of each class, and Cnt denotes sample count. Musical instru. abbreviates Musical instrument, and Falling water denotes Falling water/Rain.}
\label{tab:sound_distribution_compact}
\end{table}

\textbf{Sound event annotation:} Many Freesound clips include user-provided tags, and the perception of audio events is usually based on human hearing. Therefore, each recording is manually reviewed by listening to the audio track and assigning the heard event labels. When a label is uncertain, the recording is replayed and re-checked until a decision can be made. After manual annotation, the labels are cross-validated with the user-provided tags on Freesound.org \cite{fonseca2017freesound}. Recordings with disagreements are re-examined and corrected, and when needed, the corresponding GPS metadata is used to extract POI-derived OSM annotations as an auxiliary cue to support label verification. After review, each audio clip is paired with its POI-derived OSM annotations to form an Audio–GSC pair in Geo-ATBench. The initial annotation took about 600 person-hours, and cross-validation and re-checking took about 200 additional person-hours, for a total of about 800 person-hours over four months. End-to-end dataset collection, preparation, and annotation took about six months.

\begin{figure}[t]
\centering
 \setlength{\abovecaptionskip}{0.1cm}    
	\setlength{\belowcaptionskip}{-0.3cm} 
\includegraphics[width=0.95\columnwidth]{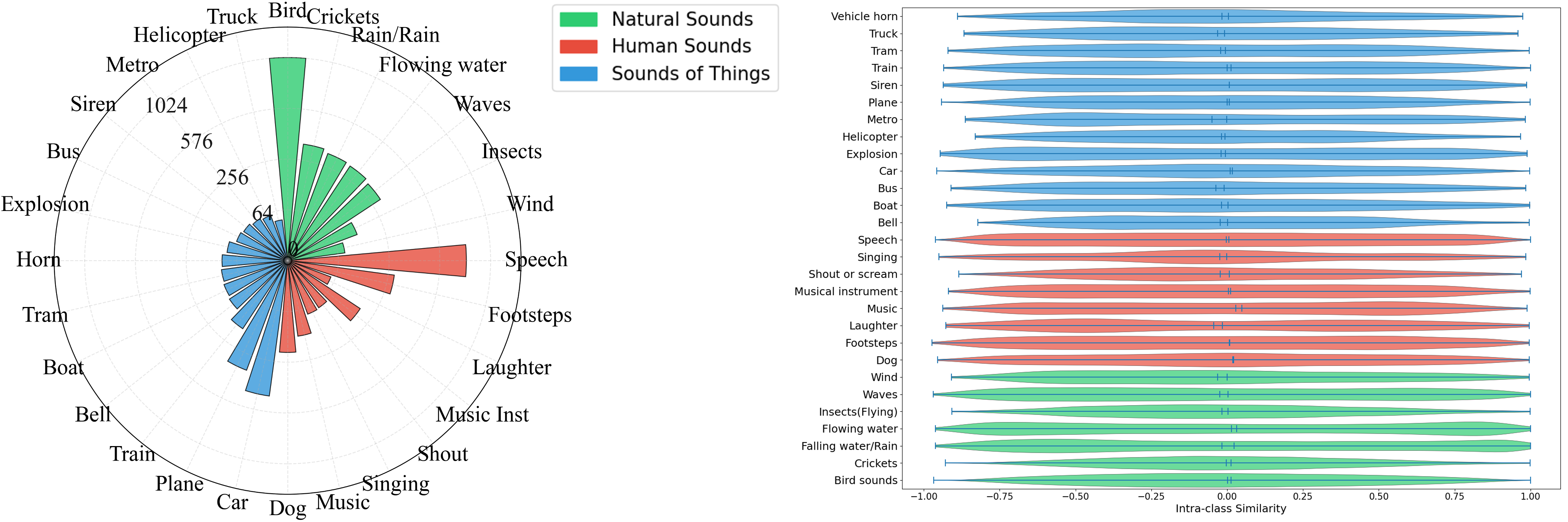}
\caption{Summary of sound classes and acoustic similarity. (Left) Distribution of three coarse-grained sound classes. (Right) Intra-class similarity across 28 sound event classes computed from log-Mel spectrogram features.}
\label{fig:combined_figure}
\end{figure}

\begin{figure}[t]
    \centering
     \setlength{\abovecaptionskip}{0.05cm}    
	\setlength{\belowcaptionskip}{-0.3cm} 
    \includegraphics[width=0.8\textwidth]{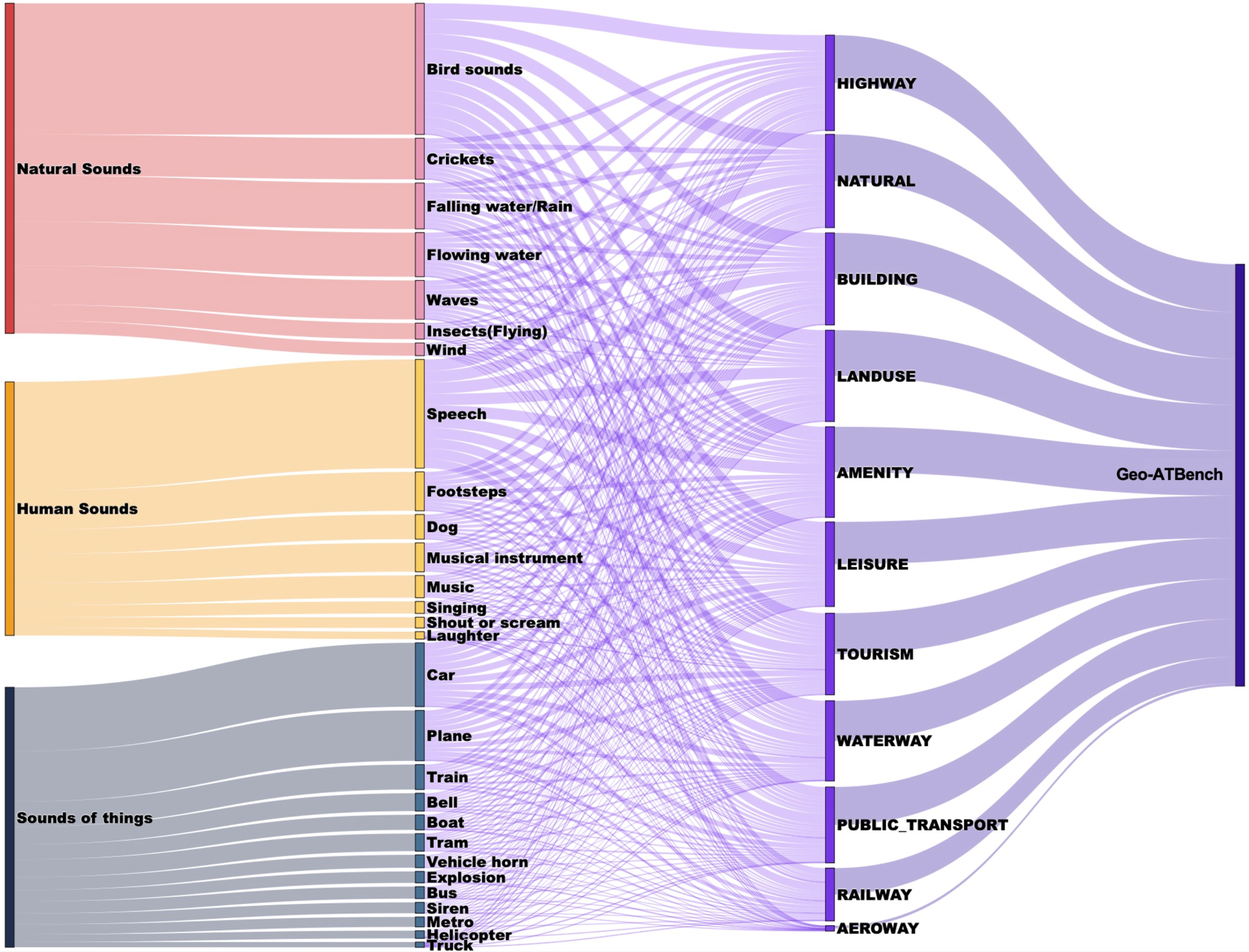}
    \caption{Sankey diagram summarizing co-occurrence links from left to right: 3 coarse-grained sound classes, 28 fine-grained sound event classes, GSC types, and the Geo-ATBench dataset. Flow width indicates co-occurrence strength. This diagram represents the distribution of audio events and GSC types within the dataset, and is not intended to imply precise real-world relationships, as sound occurrences can vary significantly depending on the specific geographical context (e.g., residential roads vs highways).}
    \label{fig:poi_distribution}
\end{figure}

A curation process is performed to map unstructured annotation labels into a controlled vocabulary, resulting in 28 sound event classes. These 28 classes are grouped into three main categories aligned with the AudioSet taxonomic structure~\cite{Gemmeke2017audioset}: 1) \textbf{Natural Sounds}, which include sounds originating from nature; 2) \textbf{Human Sounds}, which encompass sounds produced by humans; and 3) \textbf{Sounds of Things}, which represent mechanical and man-made noises. The sample counts and total durations for these categories are illustrated in Table~\ref{tab:sound_distribution_compact}, while the coarse-grained distributions and corresponding intra-class similarities are visualized in Fig.~\ref{fig:combined_figure} (right), where the violin-plot similarities are calculated based on log-Mel spectrogram features, and similarity is measured using cosine similarity between feature vectors, a widely adopted metric in audio and sound analysis~\cite{schutze2008introduction}. Additionally, Fig.~\ref{fig:poi_distribution} provides an overview of the dataset's composition, encompassing 28 event types and 11 OSM categories. The dataset is inherently multi-label, accounting for the co-occurrence of multiple sound events within a single 10-second recording.

\subsection{Dataset Organization and Statistics} 

Following cleaning and selection, the final Geo-ATBench dataset comprises 3,854 audio clips, totaling 10.71 hours of audio. Each data point consists of a triplet: (i) a 10-second audio clip, (ii) a multi-label clip-level label vector over 28 event classes, and (iii) a POI-derived GSC representation constructed from OSM annotations over 11 semantic context categories. 
To ensure consistency for modeling tasks, all collected recordings are processed into a standardized format. Each audio clip has a fixed duration of 10 seconds, encoded as a single-channel (mono) WAV file with a sampling rate of 16 kHz and a bit depth of 16. For more details and access to the dataset, please visit the project homepage.

\section{The GeoFusion-AT framework and instantiations}\label{geo-at-fusion}

\subsection{The GeoFusion-AT framework}\label{geo-framework}

\begin{figure}[t]
    \centering
        \setlength{\abovecaptionskip}{0.05cm}    
	\setlength{\belowcaptionskip}{-0.3cm} 
    \includegraphics[width=0.8\textwidth]{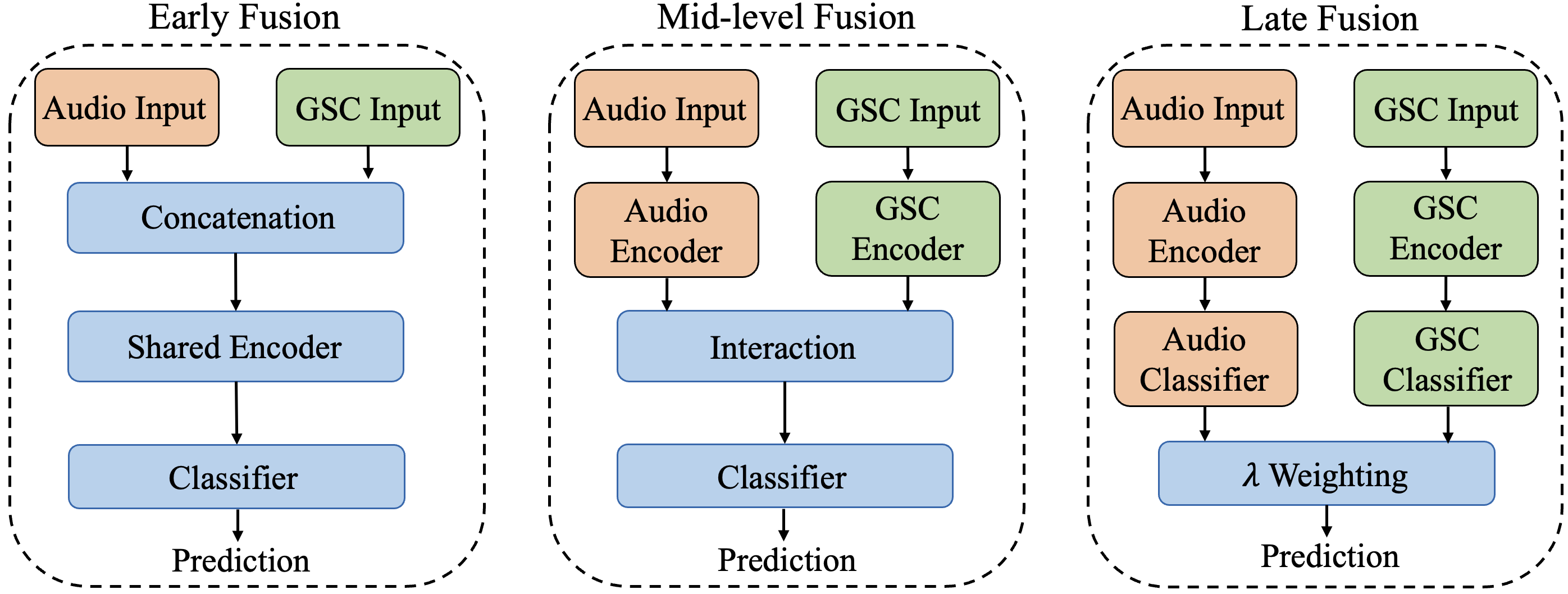}
    \caption{Overall architecture of the GeoFusion-AT framework for Geo-AT task.}
    \label{fig:simple_net}
\end{figure}

As shown in Fig.~\ref{fig:simple_net}, GeoFusion-AT provides reference implementations of three typical fusion points for the Geo-AT task on the Geo-ATBench dataset. All variants take paired inputs $(\mathbf{A}, \mathbf{g})$ and output multi-label logits $\mathbf{z}\in\mathbb{R}^{C}$ for $C$ event classes, followed by a sigmoid for tag probabilities.
 
\subsubsection{GeoFusion-Early: feature-level fusion}
\label{ssec:early_fusion}
 
Early fusion~\cite{snoek2005early}, also known as feature-level fusion, integrates geospatial context and acoustic information at the input of the network. The process begins by transforming the raw audio waveform into a log-Mel spectrogram $\mathbf{A} \in \mathbb{R}^{1 \times T \times F}$, where $T$ and $F$ denote the number of time frames and frequency bins, respectively. Concurrently, GSC vector $\mathbf{g} \in \mathbb{R}^{D_{\text{GSC}}}$ is projected into a length-$F$ vector $\mathbf{g}'\in \mathbb{R}^{F}$ via a linear transformation: $\mathbf{g}' = \mathbf{W}_{\text{proj}} \mathbf{g}$,
where $\mathbf{W}_{\text{proj}} \in \mathbb{R}^{F \times D_{\text{GSC}}}$ is a learnable projection matrix. We choose the frequency resolution $F$ for the projection so that $\mathbf{g}'$ can be interpreted as a location-conditioned spectral prior (i.e., a per-frequency weighting/gating signal): different geographic contexts tend to correlate with different dominant sound sources and background noise, which manifest as characteristic energy distributions over frequency bands. The projected vector $\mathbf{g}'$ is then broadcast across the temporal dimension to form a broadcast GSC tensor $\mathbf{G} \in \mathbb{R}^{1 \times T \times F}$.   

The audio spectrogram and the broadcast GSC tensor are concatenated along the channel dimension to produce the fused representation $\mathbf{X}_{\text{fused}} = \operatorname{Concat}(\mathbf{A}, \mathbf{G}) \in \mathbb{R}^{2 \times T \times F}$, which serves as the input to the backbone network. When a backbone does not accept a two-channel spectrogram input, an input adapter is applied to map $\mathbf{X}_{\text{fused}}$ into the backbone’s expected input shape and channel format; all subsequent backbone components remain unchanged.

\subsubsection{GeoFusion-Inter: representation-level fusion}
\label{ssec:inter_fusion}

Intermediate fusion~\cite{ngiam2011multimodal}, or representation-level fusion, combines information in the latent space after each modality has been processed by separate encoders. Let $\Phi_{\text{audio}}$ be an audio encoder that maps an input spectrogram $\mathbf{A}$ to an audio embedding $\mathbf{E}_{\text{audio}} \in \mathbb{R}^{D_{\text{emb}}}$, where $D_{\text{emb}}$ is the embedding dimension. Similarly, the GSC vector $\mathbf{g}$ is processed through a multi-layer perceptron (MLP) projection to produce a GSC embedding $\mathbf{E}_{\text{GSC}} \in \mathbb{R}^{D_{\text{emb}}}$ of the same dimension. Here, both embeddings are clip-level representations, implying that temporal information in $\mathbf{A}$ has been aggregated by $\Phi_{\text{audio}}$ prior to fusion.

Intermediate fusion implements a symmetric cross-modal attention \cite{akbari2021vatt} module that supports bidirectional refinement between the audio and GSC embeddings. Given $\mathbf{Q}, \mathbf{K}, \mathbf{V}$ are the query, key and value, attention is computed as $\text{Attention}(\mathbf{Q}, \mathbf{K}, \mathbf{V}) = \text{softmax}\left(\frac{\mathbf{Q}\mathbf{K}^T}{\sqrt{D_{\text{emb}}}}\right)\mathbf{V}$, where $\mathbf{K}=\mathbf{V}$, the factor $\sqrt{D_{\text{emb}}}$ stabilizes optimization \cite{akbari2021vatt}. Accordingly, the cross-modal attention operates on global embeddings rather than temporal tokens, serving as feature-wise conditioning instead of frame-level alignment. The audio embedding $\mathbf{E}_{\text{audio}}$ is enhanced by treating it as a query and the GSC embedding $\mathbf{E}_{\text{GSC}}$ as the key/value. Symmetrically, $\mathbf{E}_{\text{GSC}}$ is enhanced using $\mathbf{E}_{\text{audio}}$ as context. Two complementary fusion streams are formed by residual mixing. One stream combines the cross-attention refined audio embedding with the original GSC embedding. The other stream combines the cross-attention refined GSC embedding with the original audio embedding. This symmetric design preserves both the cross-modal updates and the original modality information. The two streams are then concatenated and passed through a learnable linear projection to produce a single fused embedding, which is fed to the classification head to output multi-label tagging logits.

\subsubsection{GeoFusion-Late: decision-level fusion}
\label{ssec:late_fusion}

Late fusion~\cite{snoek2005early}, or decision-level fusion, combines the outputs of two independent streams, one for each modality. In this paradigm, an audio branch, $\Phi_{\text{audio}}$, processes the audio representation $\mathbf{A}$ to produce class-wise logits, $\mathbf{z}_{\text{audio}} \in \mathbb{R}^{C}$, where $C$ is the number of event classes. In parallel, a GSC branch, $\Phi_{\text{GSC}}$, takes the POI-derived GSC vector $\mathbf{g}$ as input and produces its own logits, $\mathbf{z}_{\text{GSC}}$. The fusion is performed by a weighted combination of these two logits. Rather than using a single scalar weight, a learnable, class-specific weighting vector $\boldsymbol{\lambda} \in \mathbb{R}^{C}$ is used. This design assigns a separate GSC weight to each class while keeping the audio branch unchanged. The fused logits $\mathbf{z}_{\text{fused}}$ are computed as:
\begin{equation}
\setlength{\abovedisplayskip}{-1pt}
\setlength{\belowdisplayskip}{-1pt} 
    \mathbf{z}_{\text{fused}} = \mathbf{z}_{\text{audio}} + \boldsymbol{\lambda} \odot \mathbf{z}_{\text{GSC}}
    \label{eq:late_fusion}
\end{equation}
where $\odot$ denotes element-wise multiplication, $\boldsymbol{\lambda}$ is constrained to be non-negative via a softplus activation function~\cite{glorot2011deep}, $\boldsymbol{\lambda} = \text{softplus}(\boldsymbol{\lambda}_{\text{raw}})$, and $\boldsymbol{\lambda}_{\text{raw}}$ is zero-initialized. 
The fusion is performed in the logit (pre-sigmoid) domain, where $\mathbf{z}$ denotes class-wise log-odds scores. Thus, Eq.~(\ref{eq:late_fusion}) combines modality-specific evidence before the final sigmoid mapping to probabilities. The final class probabilities are obtained by applying a sigmoid function to $\mathbf{z}_{\text{fused}}$.


The GeoFusion-AT framework uses the standard multi-label AT objective on the fused logits. Auxiliary losses and regularizers are optional and not required by the framework definition.

\subsection{Instantiations of the GeoFusion-AT framework} \label{section:instantiation} 
GeoFusion-AT is instantiated on three representative audio backbones to provide benchmark results for the Geo-AT task. PANNs \cite{Kong2020panns} is a CNN-based pretrained audio backbone, AST~\cite{Gong2021ast} is a patch-based Transformer backbone that applies attention over spectrogram patch embeddings, and CLAP~\cite{elizalde2023clap} is a contrastively pretrained audio–text backbone. All instantiations follow the definitions in Section~\ref{geo-framework}: feature-level fusion (GeoFusion-Early), representation-level fusion (GeoFusion-Inter), and decision-level fusion (GeoFusion-Late). Source code and model checkpoints are available on the project homepage.

\subsubsection{Instantiations of GeoFusion-Early}

\begin{figure}[t]
    \centering
        \setlength{\abovecaptionskip}{0.1cm}    
	\setlength{\belowcaptionskip}{-0.4cm} 
    \includegraphics[width=0.8\textwidth]{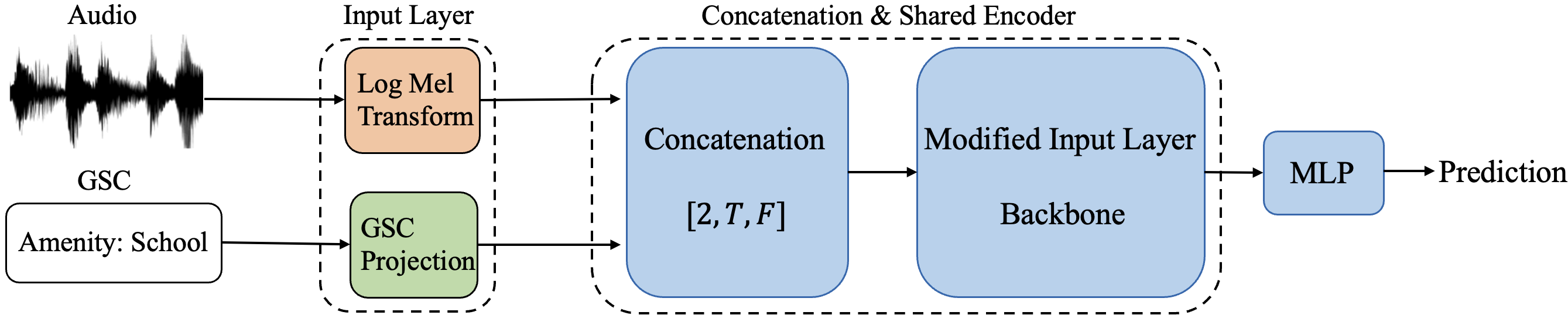}
    \caption{Instantiations of GeoFusion-Early (feature-level fusion).}
    \label{fig:earlyfusion} 
\end{figure}

GeoFusion-Early implements feature-level fusion by constructing an acoustic representation tensor and a broadcast GSC tensor, as shown in Fig.~\ref{fig:earlyfusion}.

\textbf{GeoFusion-Early-PANNs}.\label{sssec:inst_early_panns}
The instantiation on PANNs~\cite{Kong2020panns} follows Section~\ref{ssec:early_fusion}. The GSC vector $\mathbf{g} \in \mathbb{R}^{D_{\text{GSC}}}$ (with $D_{\text{GSC}} = 768$) is linearly projected to a length-$F$ vector and broadcast along time to form a broadcast GSC tensor $\mathbf{G}\in\mathbb{R}^{1\times T\times F}$. Audio preprocessing operations are applied to $\mathbf{A}$ before fusion. The fused input is $\mathbf{X}_{\text{fused}} = \operatorname{Concat}(\mathbf{A},\mathbf{G}) \in \mathbb{R}^{2\times T\times F}$. The first convolutional layer is adapted to accept two input channels. Weights for the audio channel are initialized from the PANNs checkpoint, and weights for the GSC channel are zero-initialized to preserve the pretrained audio pathway at initialization and let the model learn to use $\mathbf{g}$ during fine-tuning.
 
\textbf{GeoFusion-Early-AST}.
\label{sssec:inst_early_ast} 
For AST~\cite{Gong2021ast}, GeoFusion-Early is implemented as feature-level fusion in the token sequence. Instead of channel-wise concatenation, the GSC vector $\mathbf{g}$ is mapped to the AST embedding dimension and injected as a dedicated \texttt{[GSC]} token. The Transformer input sequence contains the standard \texttt{[CLS]} token, the \texttt{[GSC]} token, and the audio patch tokens. The positional embedding table is expanded to $(1, N_{\text{patches}} + 2, D_{\text{emb}})$ (with $D_{\text{emb}} = 768$), and the new \texttt{[GSC]} position is zero-initialized while the original positions retain their pretrained values from the AST checkpoint. Classification uses the output embedding of the \texttt{[CLS]} token.

\textbf{GeoFusion-Early-CLAP}.
\label{sssec:inst_early_clap} 
The CLAP audio encoder~\cite{elizalde2023clap} accepts a spectrogram input and is instantiated with the same two-channel construction as GeoFusion-Early-PANNs. A broadcast GSC tensor $\mathbf{G}$ is constructed from $\mathbf{g}$ and concatenated with $\mathbf{A}$ to form $\mathbf{X}_{\text{fused}}$. 
Weights for the audio channel are initialized from the checkpoint, while the GSC channel is zero-initialized to avoid perturbing pretrained audio representations early in training.

\subsubsection{Instantiations of GeoFusion-Inter}

\begin{figure}[b]
    \centering
        \setlength{\abovecaptionskip}{0.1cm}    
	\setlength{\belowcaptionskip}{-0.4cm} 
    \includegraphics[width=0.9\textwidth]{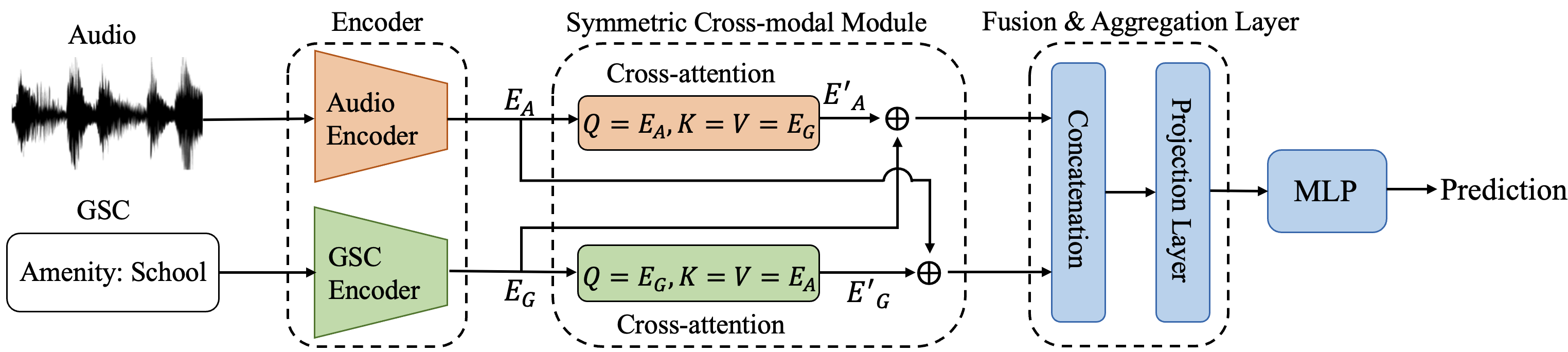}
    \caption{Instantiations of GeoFusion-Inter (representation-level fusion).}
    \label{fig:midfusion}
\end{figure}

GeoFusion-Inter is a representation-level fusion variant that combines the audio embedding $\mathbf{E}_{\text{audio}}$ and the GSC embedding $\mathbf{E}_{\text{GSC}}$ using the symmetric cross-modal attention module in Section~\ref{ssec:inter_fusion}, as shown in Fig.~\ref{fig:midfusion}.

\textbf{GeoFusion-Inter-PANNs}.
\label{sssec:inst_inter_panns}
For PANNs, its pretrained backbone serves as a feature extractor to produce audio embedding $\mathbf{E}_{\text{audio}} \in \mathbb{R}^{D_{\text{emb}}}$ ($D_{\text{emb}} = 2048$ for PANNs). In parallel, the GSC vector $\mathbf{g} \in \mathbb{R}^{D_{\text{GSC}}}$ is projected by a two-layer MLP into GSC embedding $\mathbf{E}_{\text{GSC}} \in \mathbb{R}^{D_{\text{emb}}}$. The embeddings are combined using the symmetric cross-modal attention module in Section~\ref{ssec:inter_fusion} to produce $\mathbf{E}_{\text{fused}}$, which is fed to the classification head to output multi-label logits.

\textbf{GeoFusion-Inter-AST}.
\label{sssec:inst_inter_ast}
For AST, the \texttt{[CLS]} output embedding is used as $\mathbf{E}_{\text{audio}} \in \mathbb{R}^{D_{\text{emb}}}$ with $D_{\text{emb}} = 768$. The GSC vector $\mathbf{g} \in \mathbb{R}^{D_{\text{GSC}}}$ has $D_{\text{GSC}} = 768$ and is used to form $\mathbf{E}_{\text{GSC}}$ at the same dimension. The attention module in Section~\ref{ssec:inter_fusion} produces $\mathbf{E}_{\text{fused}}$ for tagging.

\textbf{GeoFusion-Inter-CLAP}.
\label{sssec:inst_inter_clap}
For CLAP, its audio encoder produces $\mathbf{E}_{\text{audio}} \in \mathbb{R}^{D_{\text{emb}}}$ with $D_{\text{emb}} = 1024$. Concurrently, a two-layer MLP projects the GSC vector $\mathbf{g}$ into a matching GSC embedding $\mathbf{E}_{\text{GSC}}$. The attention module in Section~\ref{ssec:inter_fusion} combines the embeddings to produce $\mathbf{E}_{\text{fused}}$ for tagging.

\subsubsection{Instantiations of GeoFusion-Late}

\begin{figure}[hb]
    \centering
        \setlength{\abovecaptionskip}{0.1cm}    
	\setlength{\belowcaptionskip}{-0.3cm} 
    \includegraphics[width=0.9\textwidth]{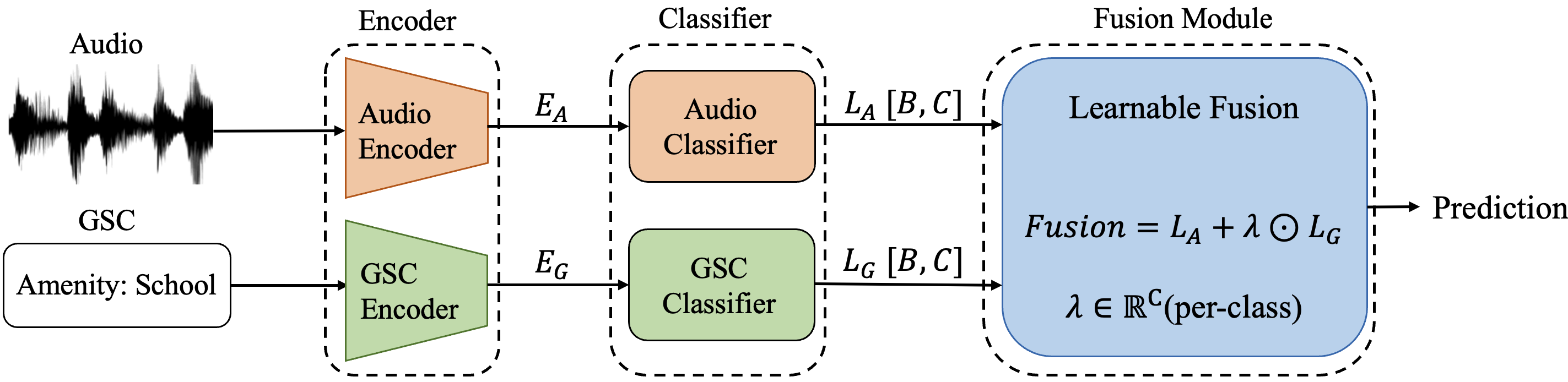}
    \caption{Instantiations of GeoFusion-Late (decision-level fusion).}
    \label{fig:latefusion}
\end{figure}

GeoFusion-Late implements decision-level fusion by combining audio logits $\mathbf{z}_{\text{audio}}$ and GSC logits $\mathbf{z}_{\text{GSC}}$ using Eq.~\ref{eq:late_fusion}, as shown in Fig.~\ref{fig:latefusion}.

\textbf{GeoFusion-Late-PANNs}.
\label{sssec:inst_late_panns}
The audio branch is the PANNs model and outputs $\mathbf{z}_{\text{audio}}$. The GSC branch is an MLP that maps $\mathbf{g}$ to $\mathbf{z}_{\text{GSC}}$. Fused logits $\mathbf{z}_{\text{fused}}$ are computed via Eq.~\ref{eq:late_fusion} and optimized with the multi-label objective.

\textbf{GeoFusion-Late-AST}.
\label{sssec:inst_late_ast}
The audio branch is the AST model and outputs $\mathbf{z}_{\text{audio}}$. The GSC branch and logit fusion follow GeoFusion-Late-PANNs, producing $\mathbf{z}_{\text{fused}}$ for multi-label tagging.

\textbf{GeoFusion-Late-CLAP}.
\label{sssec:inst_late_clap}
The audio branch uses the CLAP audio encoder to produce an audio embedding, followed by a classification head to output $\mathbf{z}_{\text{audio}}$. The GSC branch and logit fusion follow GeoFusion-Late-PANNs.

\section{Experiments and results analysis}

\subsection{Experimental setup and evaluation metrics}  

Geo-ATBench is evaluated as a 28-class multi-label AT task; each audio clip is represented by the acoustic input and the paired POI-derived GSC, described in Section~\ref{section_task} and Section~\ref{section_dataset}. For repeated evaluation, five independent runs are conducted with different random seeds. In each run, the dataset is split into 70\% training, 15\% validation, and 15\% test. A multi-label stratification procedure is used to keep per-label prevalence and co-occurrence patterns comparable across splits so that all event classes are represented in the test set. The split is performed at the clip level. The GSC representation is not constructed from precise geographic identifiers such as GPS coordinates, street names, or addresses. Instead, it encodes high-level semantic context derived from nearby POIs. Specifically, raw OSM tags, such as amenity: school and highway: bus stop, are extracted and converted into descriptive strings. The resulting strings are encoded using a pretrained BERT model~\cite{devlin2019bert}, and element-wise mean pooling is applied to the embeddings to capture local land-use characteristics and area semantics. Similar GSC patterns may occur across different recording locations, while recordings in the same area may still differ in their POI composition. Thus, the reported benchmark results should be interpreted as evaluating generalization under clip-level partitioning with location-derived semantic context, rather than under strict geographic hold-out.

The three backbones (PANNs \cite{Kong2020panns}, AST \cite{Gong2021ast}, and CLAP \cite{elizalde2023clap}) used in this paper are pretrained on large-scale AudioSet \cite{Gemmeke2017audioset} and have reported strong performance on AudioSet with 527 audio event classes at the time of their proposal. In the benchmark construction for Geo-AT on Geo-ATBench, fine-tuning is applied to adapt these backbones to the 28-class multi-label task while limiting changes to their pretrained audio representations. This is enforced through a small learning rate and early stopping. 
Models are trained on an NVIDIA GeForce RTX 4090 GPU and fine-tuned for a maximum of 100 epochs using the AdamW optimizer with a learning rate of 1e-5. Early stopping is applied to prevent overfitting; training stops if the validation F1 score does not improve for 15 consecutive epochs. The training objective is binary cross-entropy (BCE) loss~\cite{Kong2020panns}. Audio inputs are 10-second clips and are resampled to match each backbone’s requirements. All models are initialized from pretrained weights, and audio-only baselines are included for comparison.

Model performance is evaluated by mean Average Precision (mAP) \cite{Kong2020panns}, area under the ROC curve (ROC AUC), 
and F1 score, with mean $\pm$ standard deviation across the 5 independent runs. All metrics are micro-averaged unless stated otherwise. Besides the multi-label AT on 28 event classes, a 3-class coarse-grained AT is reported as a supplementary analysis. For all source code, models, and the dataset, please see the project homepage.

\subsection{Results and analysis}

This section evaluates Geo-ATBench from three complementary perspectives. First, the feasibility of performing audio tagging with GSC alone is evaluated as a GSC-only baseline under different POI extraction ranges. Second, audio-only zero-shot baselines are reported for three strong AudioSet-pretrained audio backbones to characterize backbone behaviour before fine-tuning on Geo-ATBench. Third, fine-tuned Geo-AT results on Geo-ATBench are reported for audio-only and GeoFusion-AT variants under identical data splits, enabling a controlled comparison of feature-level, representation-level, and decision-level fusion. Per-label performance changes and error patterns are used to identify which labels and confusions benefit most from GSC, with emphasis on acoustically confusable labels.

\subsubsection{GSC-only baseline and GSC range sensitivity} 

\begin{figure}[tb]
    \centering
        \setlength{\abovecaptionskip}{0.05cm}    
	\setlength{\belowcaptionskip}{-0.5cm} 
    \includegraphics[width=\textwidth]{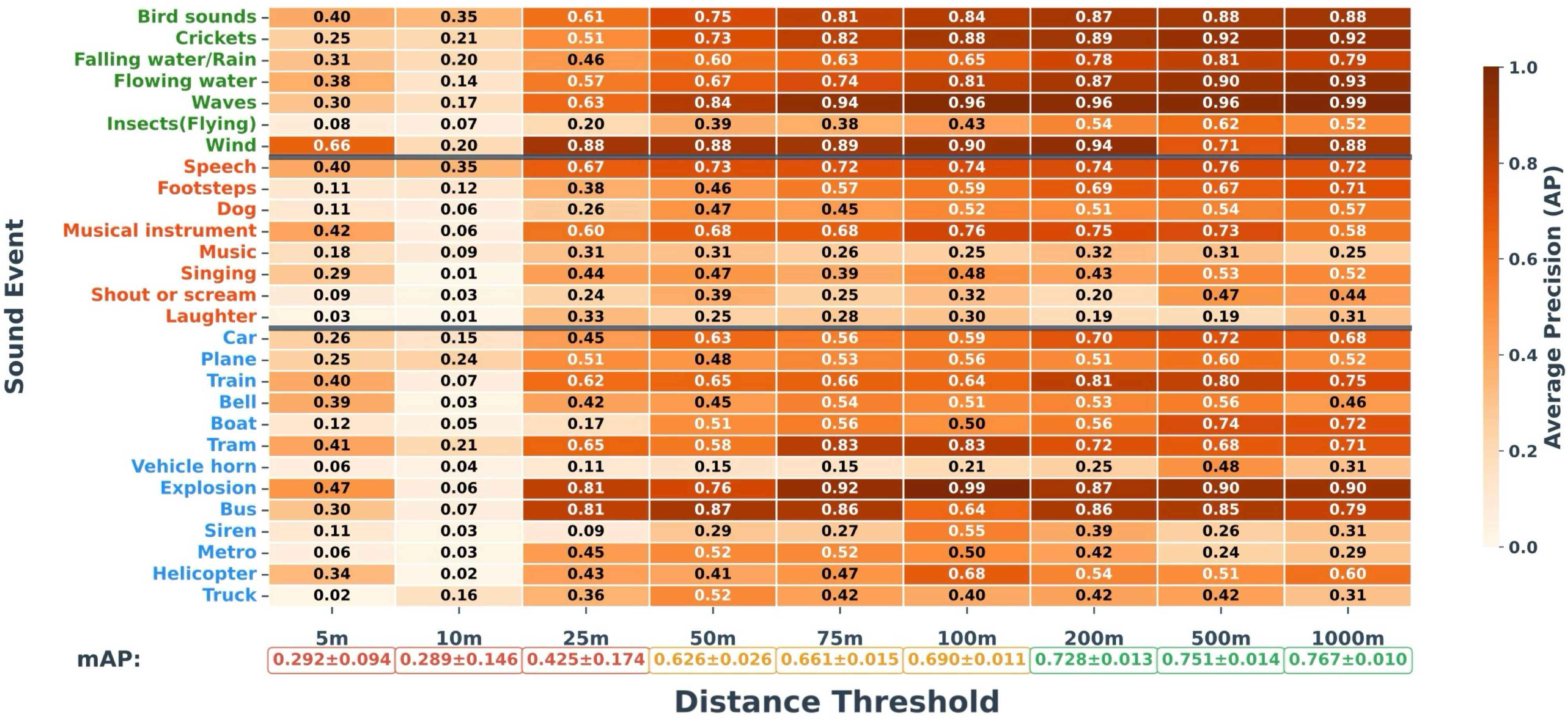}
    \caption{Average Precision (AP) for GSC-only multi-label tagging under different POI extraction ranges. 
    POI-derived GSC is constructed from OSM entities retrieved with the same 11 OSM feature keys (e.g., land use, amenities) and encoded into the fixed-length GSC vector $\mathbf{g}$. mAP is computed on the test set over 5 independent runs.}
    \label{fig:cat_poi_distribution}
\end{figure}

In practice, sound events differ in how broadly they can be perceived and in how strongly they correlate with nearby place semantics. As a result, the POI extraction range affects the amount and composition of POI-derived context available for constructing the GSC vector $\mathbf{g}$. To benchmark the Geo-only baseline on Geo-ATBench, a GSC-only baseline is evaluated under multiple POI extraction ranges, as shown in Fig.~\ref{fig:cat_poi_distribution}.

For each POI extraction range defined by a distance threshold, implemented as the square side length, a square is centered at the clip’s GPS coordinate. Although a circular neighborhood may better approximate isotropic sound propagation, a square region is employed to enable efficient bounding-box queries in OSM-based geographic information systems. OSM entities within the square are retrieved using the same 11 OSM feature keys \cite{OSM} as in Section~\ref{section_dataset}. The resulting POI composition is encoded by a pretrained BERT model~\cite{devlin2019bert} into the fixed-length 768-dimensional GSC vector $\mathbf{g}$, and the same GSC-only classifier is evaluated across all ranges. The Geo-only baseline uses BERT-base to produce $\mathbf{g}$, followed by a 3-layer MLP with 1024, 512, and 28 units to perform 28-label multi-label tagging on Geo-ATBench. During training, the BERT tokenizer and BERT encoder~\cite{devlin2019bert} are frozen, and only the 3-layer MLP classifier is fine-tuned. Source code, extracted GSC vectors, and implementation details are available on the project homepage.

The GSC-only results increase with larger distance thresholds on Geo-ATBench, and the 1000-metre range yields the highest performance. One possible explanation is that OpenStreetMap (OSM) coverage can be sparse in some regions \cite{herfort2023spatio}, so smaller squares may return fewer entities for constructing the GSC vector $\mathbf{g}$. GPS accuracy can also vary across devices and conditions~\cite{zandbergen2009accuracy}, which can shift the queried area and affect POI retrieval. Additional factors may also contribute, as sound events differ in how broadly their semantics relate to nearby places and in how tightly they align with POI-derived context. For example, mobile sources such as birds or crickets can be heard across a natural area and may be associated with woodland or water POIs beyond the immediate vicinity of the recording point. Human speech can also reflect nearby attractions or pedestrian flow, where people move toward or away from a site and produce speech sounds over a broader area. In contrast, events associated with fixed sources, such as breaking waves at a shoreline, are typically constrained to more local place semantics; thus, shorter ranges can be sufficient in some cases. In summary, this section presents Geo-only performance on Geo-ATBench with different POI extraction ranges, providing a detailed reference for Geo-only comparison on the proposed Geo-ATBench dataset.

\subsubsection{Audio-only zero-shot baselines}
 
Audio-only zero-shot tagging inference is reported to characterize three AudioSet-pretrained models' behaviour before fine-tuning on Geo-ATBench. The AudioSet-pretrained backbones PANNs~\cite{Kong2020panns}, AST~\cite{Gong2021ast}, and CLAP~\cite{elizalde2023clap} are evaluated through direct inference, providing a reference point for the fine-tuned audio-only and GeoFusion-AT results reported later.
 
A direct zero-shot benchmark on Geo-ATBench labels is not possible from the original AudioSet-pretrained model outputs, because these backbones are trained on AudioSet with 527-class event labels, and their native outputs do not match the 28 Geo-ATBench event labels. A comparable 28-label zero-shot benchmark is defined by first producing class predictions over the 527 AudioSet labels for each Geo-ATBench clip, and then mapping these 527 outputs to the 28 Geo-ATBench labels using the pretrained Word2Vec model (“word2vec-google-news-300”) \cite{word2vec}, which provides 300-dimensional word embeddings trained on the Google News corpus. The AudioSet-to-Geo-ATBench label mapping and code are released on the project homepage for reproduction.

\begin{figure}[t]
    \centering
        \setlength{\abovecaptionskip}{0.05cm}    
	\setlength{\belowcaptionskip}{-0.5cm} 
    \includegraphics[width=\columnwidth]{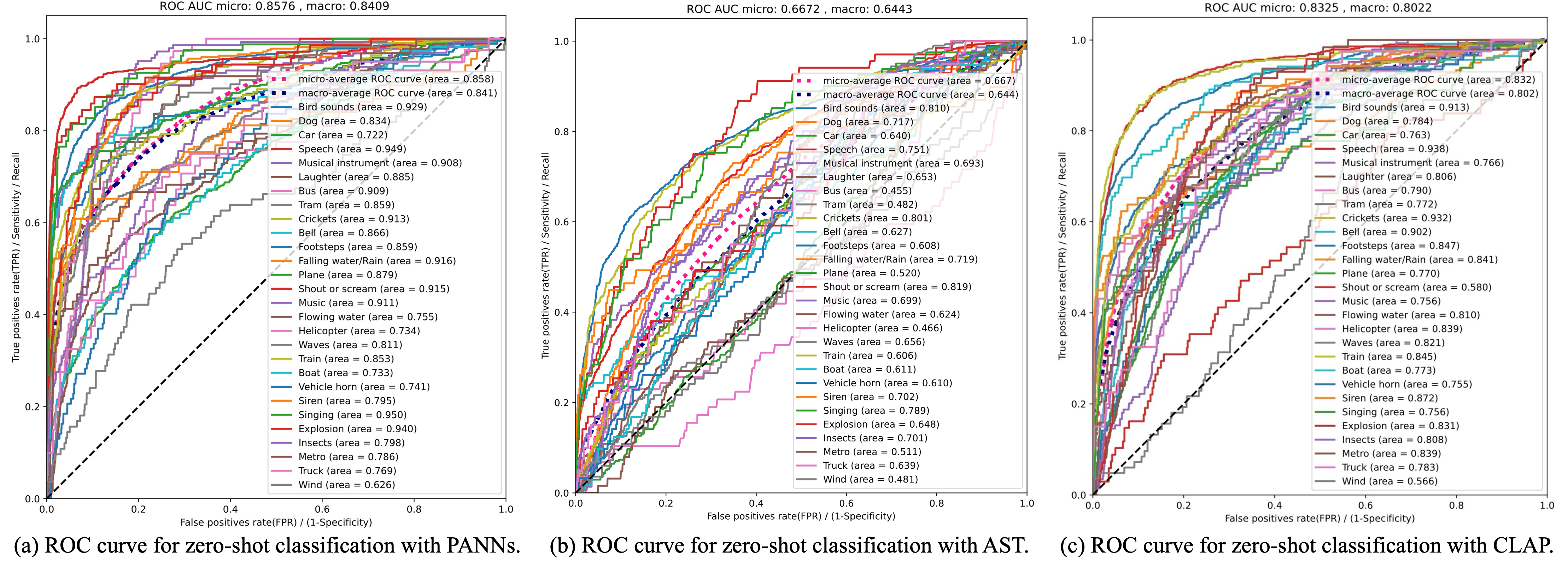}
    \caption{ROC curves for zero-shot audio-only tagging inference on Geo-ATBench labels. Micro and macro ROC AUC are reported for AudioSet-pretrained PANNs, AST, and CLAP after AudioSet-to-Geo-ATBench label mapping. Overall performance: PANNs (Micro AUC 0.8576, Macro AUC 0.8409), AST (0.6672, 0.6443), and CLAP (0.8325, 0.8022).}
    \label{fig:result1}
\end{figure}

\begin{figure}[t]
    \centering
        \setlength{\abovecaptionskip}{0.05cm}    
	\setlength{\belowcaptionskip}{-0.4cm} 
    \includegraphics[width=\columnwidth]{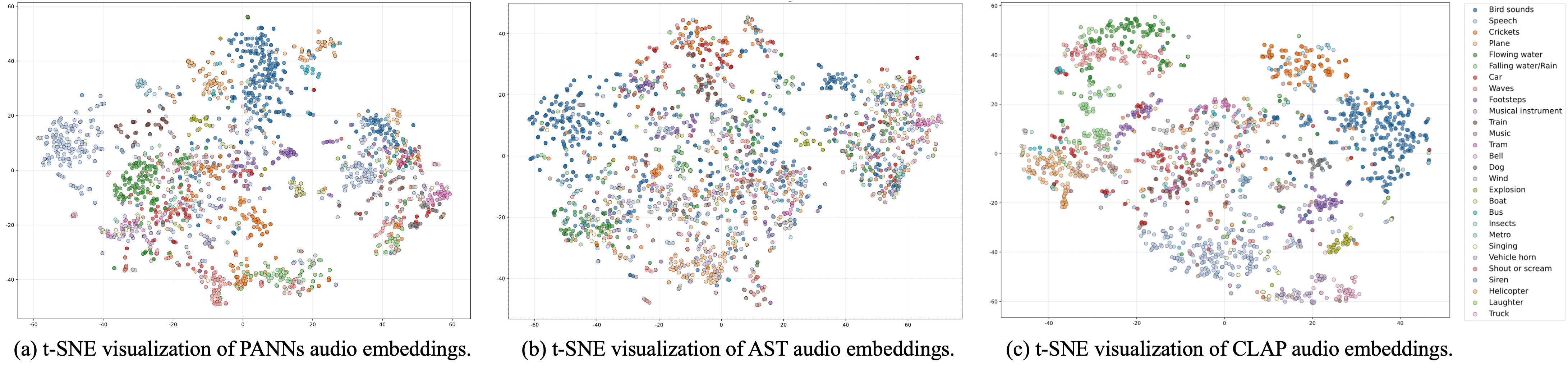}
    \caption{t-SNE visualization of audio embeddings from zero-shot inference for PANNs, AST, and CLAP on the 28 Geo-ATBench event classes; clusters show effective separation, while overlaps highlight acoustically similar classes.} 
    \label{fig:tsne}
\end{figure}

Fig.~\ref{fig:result1} shows zero-shot audio-only tagging performance on Geo-ATBench for three AudioSet-pretrained backbones. PANNs achieves the strongest performance, followed by CLAP, while AST performs the worst under the same AudioSet-to-Geo-ATBench label mapping. Several factors may contribute to this ordering. First, PANNs is trained to produce strong clip-level tag predictions from log-Mel inputs \cite{Kong2020panns}, which can transfer more directly to Geo-ATBench under label-space mapping. Second, CLAP learns aligned audio-text representations \cite{elizalde2023clap}, which can preserve semantic separation that remains useful after mapping AudioSet labels to Geo-ATBench labels. Third, AST relies on spectrogram patch tokenization and positional embeddings \cite{Gong2021ast}, and its AudioSet pre-training configuration may transfer less effectively to the Geo-ATBench distribution under direct inference without task-specific adaptation. Similarly, the visualisation of the audio embeddings under zero-shot inference shows the same trend in Fig.~\ref{fig:tsne}. Embeddings from PANNs and CLAP form more separable clusters across the 28 Geo-ATBench classes, whereas AST embeddings show stronger overlap and concentrate in a smaller region of the space. Higher-resolution versions of Fig.~\ref{fig:result1} and Fig.~\ref{fig:tsne} are available on the project homepage due to space constraints.

\vspace{-2mm}
\subsubsection{Fine-tuned Geo-AT results on Geo-ATBench}
 
Table~\ref{tab:all_result} shows the results of audio backbones under the three fusion strategies described in Section~\ref{geo-at-fusion}. Audio-only and GSC-only baselines are reported, and GeoFusion-AT variants are compared under identical data splits on the 28-class Geo-AT task. All fine-tuned models are trained on the 28 Geo-ATBench labels, without the AudioSet label mapping used in the zero-shot baselines.

\begin{table}[b]
\centering
\small
	\setlength{\abovecaptionskip}{0.1cm}   
	\setlength{\belowcaptionskip}{-0.2cm}  
\setlength{\tabcolsep}{3.5pt}
\resizebox{\textwidth}{!}{%
    \begin{tabular}{l|ccc|ccc} 
    \toprule
    \multirow{2}{*}{\textbf{Strategy}} & \multicolumn{3}{c|}{\textbf{Fine-grained (28 classes)}} & \multicolumn{3}{c}{\textbf{Coarse-grained (3 classes)}} \\
    \cmidrule(lr){2-4} \cmidrule(lr){5-7}
     & PANNs & AST & CLAP & PANNs & AST & CLAP \\
    \midrule
    Audio-Only & 0.770±0.006 & 0.820±0.015 & 0.824±0.008 & 0.961±0.006 & 0.904±0.012 & 0.966±0.005 \\
    GSC-Only & \multicolumn{3}{c|}{0.767±0.010} & \multicolumn{3}{c}{0.867±0.009} \\
    \midrule
    GeoFusion-Early & 0.812±0.010 & 0.846±0.010 & 0.826±0.010 & 0.954±0.004 & 0.914±0.009 & 0.950±0.008 \\
    \hspace{1em}\textit{Gain ($\Delta$)} & +0.042 & +0.026 & +0.002 & -0.007 & +0.010 & -0.016 \\
    \midrule
    GeoFusion-Inter & 0.824±0.010 & 0.829±0.003 & 0.842±0.006 & 0.964±0.008 & 0.912±0.011 & 0.968±0.005 \\
    \hspace{1em}\textit{Gain ($\Delta$)} & +0.054 & +0.009 & +0.018 & +0.003 & +0.008 & +0.002 \\
    \midrule
    GeoFusion-Late & 0.833±0.007 & 0.843±0.010 & 0.831±0.007 & 0.949±0.004 & 0.939±0.008 & 0.966±0.004 \\
    \hspace{1em}\textit{Gain ($\Delta$)} & +0.063 & +0.023 & +0.007 & -0.012 & +0.035 & 0.000 \\
    \bottomrule
    \end{tabular}%
}  
\caption{The mean average precision (mAP) of different models on the Geo-ATBench dataset. 
The rows labeled \textit{Gain ($\Delta$)} represent the performance difference relative to the audio-only baseline. All metrics are averaged across 5 independent experimental runs.}
\label{tab:all_result}
\end{table}
 
\begin{figure}[t]
    \centering
      \setlength{\abovecaptionskip}{0.0cm}    
	\setlength{\belowcaptionskip}{-0.3cm} 
    \includegraphics[width=0.9\columnwidth]{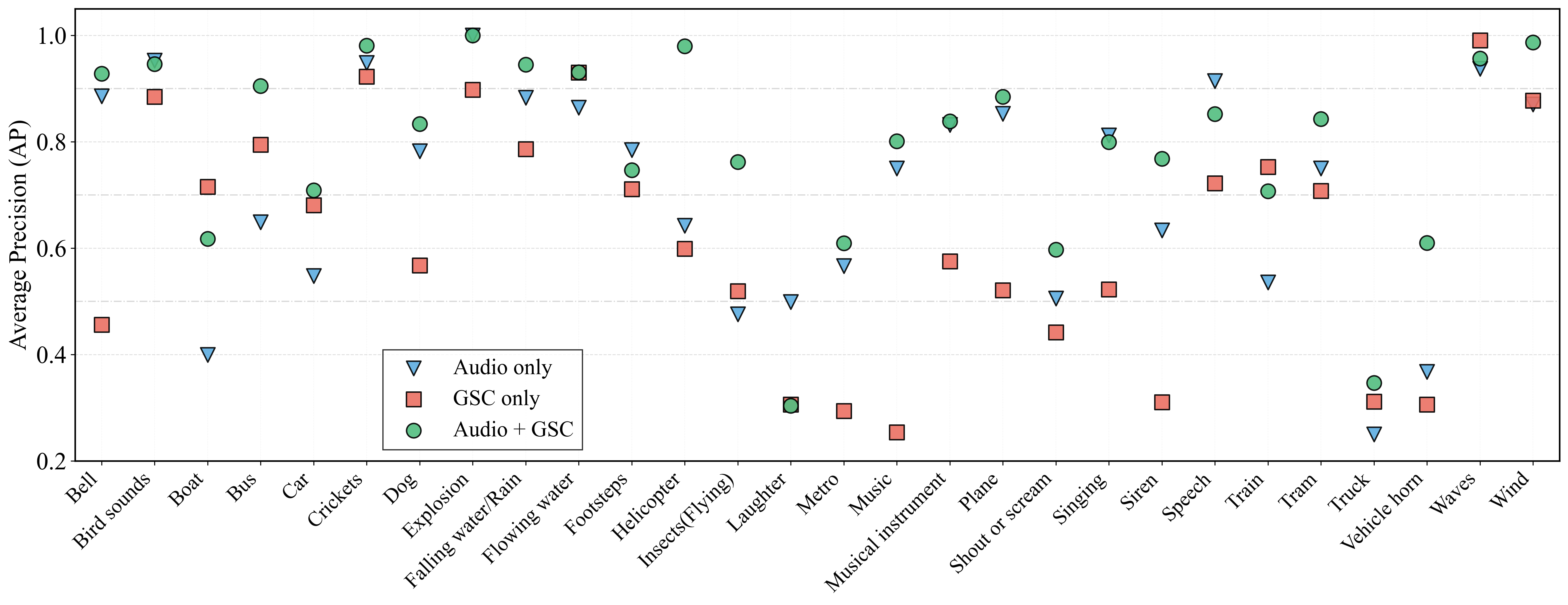}  
    \caption{Per-class average precision across 28 classes for the audio-only AST, the GSC-only baseline, and GeoFusion-Early-AST, the integration of geospatial information improves performance for multiple classes.}
    \label{fig:compare_audio_poi}
\end{figure}

Across early, representation-level, and late fusion, incorporating GSC improves 28-class Geo-AT performance for all three backbones. Over five runs, Welch two-sample t-tests indicate significant improvements compared with the corresponding audio-only baselines for AST with early fusion ($p < 0.05$), PANNs with late fusion ($p < 0.001$), and CLAP with intermediate fusion ($p < 0.01$). GeoFusion-Early-AST achieves the best mAP on the fine-grained 28-class multi-label tagging task, while no significant difference is observed between GeoFusion-Early-AST and GeoFusion-Inter-CLAP ($p > 0.5$), indicating comparable performance. After fine-tuning, AST yields the strongest overall performance on the 28-class Geo-AT task, followed by CLAP and then PANNs. This ordering differs from the zero-shot baseline ranking in Fig.~\ref{fig:result1}. The key difference is that the zero-shot baseline predicts in the 527-class AudioSet label space and then maps the outputs to the 28 Geo-ATBench labels, whereas fine-tuned models are trained directly on Geo-ATBench labels. The mapping step can introduce label-aggregation and calibration effects that vary across backbones, and cross-dataset domain shift further limits direct transfer under direct inference. Fine-tuning removes the label-space mismatch by optimizing directly on Geo-ATBench targets, resulting in better results. Fig.~\ref{fig:compare_audio_poi} further shows the average precision of GeoFusion-Early-AST across Geo-ATBench event classes.

In addition to the 28-class fine-grained tagging task, Table~\ref{tab:all_result} shows a supplementary 3-class coarse-grained tagging task that groups the 28 events into Natural Sounds, Human Sounds, and Sounds of Things, as described in Section~\ref{coarse_grained_3class}. On this coarse-grained task, GeoFusion-Inter-CLAP achieves the best mAP. Representation-level fusion improves coarse-grained performance for all three backbones, suggesting that combining audio and GSC high-level representations with symmetric cross-modal attention in the GeoFusion-Inter framework is effective at this level of semantic granularity.

\begin{figure}[t]
    \centering
        \setlength{\abovecaptionskip}{0.1cm}    
	\setlength{\belowcaptionskip}{-0.3cm}  
    \includegraphics[width=0.8\columnwidth]{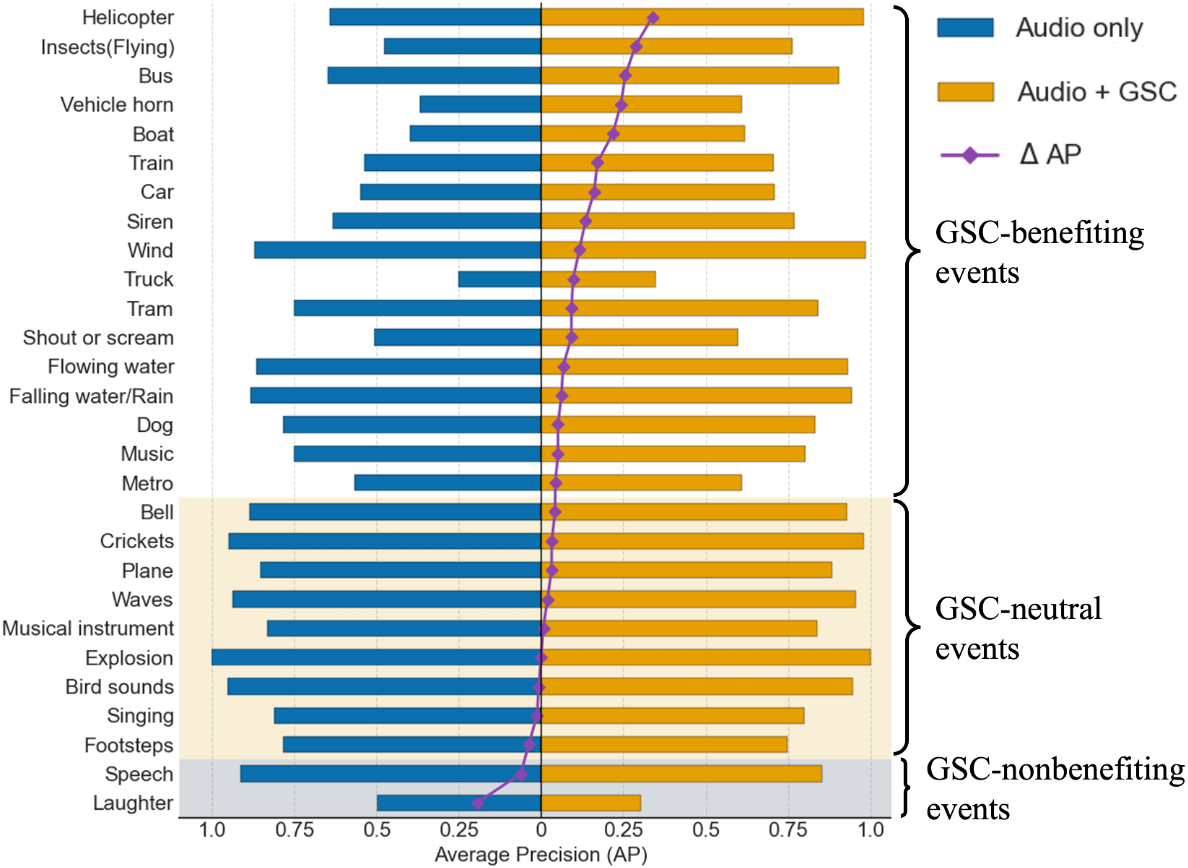}
    \caption{Per-class AP change (fusion minus audio-only) for the GeoFusion-Early-AST exemplar on the 28-class Geo-AT task.}
    \label{fig:Audio_GSC_delta_AP}
\end{figure}

To further explore which event labels benefit most from incorporating geospatial semantic context (GSC) under the Geo-AT task, Fig.~\ref{fig:Audio_GSC_delta_AP} uses GeoFusion-Early-AST as an exemplar and visualizes per-label average precision (AP) for the audio-only and audio-GSC fusion variants. Fig.~\ref{fig:Audio_GSC_delta_AP} also shows the per-label change $\Delta AP = AP_{\text{audio+GSC}} - AP_{\text{audio}}$, shown by the purple curve.  
Relative to the audio-only reference, incorporating GSC yields more than a 5\% AP increase for 17 of the 28 event classes. These 17 classes are grouped as GSC-benefiting events in this paper. Among them, \textit{Helicopter} shows the largest gain, with an absolute change of $\Delta AP = 0.3378$, corresponding to a relative increase of about $+52.62\%$ compared with the audio-only AP, which is consistent with the fact that helicopter sounds tend to occur in a limited set of places and are often associated with specific location semantics, making POI-derived GSC informative for disambiguation. 
For 9 of the 28 event classes, $\Delta AP$ remains within $\pm 5\%$, and these classes are grouped as GSC-neutral events, such as \textit{Bell}, \textit{Singing}, and \textit{Footsteps}, which are common everyday sounds and are often weakly tied to specific place semantics. It is worth noting that \textit{Explosion} shows a near-zero change in this dataset; this pattern is consistent with the Explosion samples retrieved from Freesound.org \cite{fonseca2017freesound} being dominated by daily activities such as fireworks.
Finally, two classes, \textit{Speech} and \textit{Laughter}, show decreases below $-5\%$ and are grouped as GSC-nonbenefiting events. This may be because speech and laughter are broadly distributed across locations, so associating them with POI-derived GSC does not help with recognition.
Overall, Fig.~\ref{fig:Audio_GSC_delta_AP} indicates that GSC helps for the recognition of a majority of sound event classes, has a limited impact on a subset of common sound event classes, and may not help for some widespread human vocalization events that are not related to specific locations and places.

\section{Human evaluation of the Geo-ATBench dataset} 
 
To assess how well models trained on the Geo-ATBench dataset align with human auditory judgements, a crowdsourced human listening study is conducted. This study examines the correspondence between model predictions and human multi-label event judgements. Using the collected annotations, (1) annotation agreement is summarized with descriptive consistency measures and chance-corrected reliability statistics, and (2) model–human alignment is assessed by comparing model predictions with aggregated human consensus labels at the clip level.

\subsection{Participants and Experimental Design}
 
Ten Chinese participants (3 females, 7 males; M = 22.4, SD = 0.70 years) took part in the assessment experiment. Participants shared a similar language background to support consistent understanding of the annotation interface and task instructions. The study adhered to the ethical guidelines of Xi'an Jiaotong-Liverpool University, and informed consent was obtained from all participants.
 
To assess the perceptual validity of the Geo-ATBench labels and downstream model predictions, a within-subject human annotation experiment is conducted. Participants listen to 579 Geo-ATBench audio clips and indicate event presence by selecting “exist” whenever the corresponding event is clearly heard. Multiple events may be marked as present within a single clip, consistent with the multi-label tagging formulation in Geo-AT. Each participant receives all clips and a checklist of event categories consistent with the 28 Geo-ATBench event labels. Audio clips can be replayed until a confident judgment is reached.
Participants are instructed to rely on auditory perception, consistent with the Geo-ATBench labelling procedure described in Section~\ref{coarse_grained_3class}. The annotation task is split into short sessions and presented in random order to reduce fatigue and order effects. Each participant completes the annotation of all audio clips within 14 days.

\subsection{Human listening study results: reliability and model–human alignment}
\subsubsection{Inter-rater reliability of human multi-label event annotations}
 
Descriptive agreement with the aggregated human consensus labels is computed for each participant. Across all participants, the mean agreement is 0.97, indicating that participants made similar decisions across audio clips and sound event categories. The annotation matrix is sparse, with only about 4.5\% of clip–event positions marked as 1 (present). Such class imbalance inflates raw percent agreement, because the majority of annotations belong to the same (absent) category. To obtain a chance-corrected estimate of reliability, Krippendorff's alpha for nominal data is computed. Each clip-by-event pair is treated as one item, yielding 16{,}212 items across ten participants. The resulting reliability coefficient is $\alpha_{\text{nominal}}(N = 16{,}212,\ R = 10) = 0.486$, indicating moderate agreement among 10 participants, suggesting variability in auditory perception for multi-label polyphonic sound events in individual auditory experience and interpretations. Given this moderate agreement, majority voting is used to derive clip-level consensus labels for each event as the reference for model comparison. After data collection, binary event matrices are generated for each participant and aggregated per clip–event pair: a value of 1 is assigned when at least 5 of 10 participants marked “exist”, and 0 otherwise.

Overall, the annotations show high raw agreement but only moderate chance-corrected reliability, which is expected under sparse binary multi-label tagging. Majority-vote consensus labels are used as the clip-level reference for model–human alignment, with cautious interpretation for low-prevalence events. The next subsection compares model predictions against the aggregated human consensus labels to quantify model–human alignment on Geo-ATBench.

\subsubsection{Model–human alignment under two label references}
 
To assess how sensitive model evaluation is to the choice of label reference, model predictions are evaluated against two label sets on the same test set of 579 clips: (i) the Geo-ATBench labels and (ii) the aggregated human consensus labels from 10 participants. A consensus threshold of 0.5 is used, meaning that an event is considered present when at least 5 of 10 annotators labeled it as present. This comparison aims to evaluate whether the Geo-ATBench label reference is consistent with independent human judgements.

Results are reported for the audio-only-CLAP baseline and the GeoFusion-AT variant GeoFusion-Inter-CLAP, given its strong performance on the 28-class fine-grained and 3-class coarse-grained Geo-AT tasks. Paired Wilcoxon signed-rank tests are performed on the 28 per-class F1 scores under the two label references. The result shows that for the audio-only-CLAP, there is no statistically significant difference in its performance between Geo-ATBench labels and aggregated human consensus labels, and the same conclusion applies to the GeoFusion-Inter-CLAP. Specifically, paired Wilcoxon signed-rank tests indicated that the audio-only-CLAP's F1 score does not differ significantly ($W = 181, p > 0.05$) between Geo-ATBench labels ($F1 = 0.628$) and 10 participant human consensus labels ($F1 = 0.570$). For brevity, Table~\ref{tab:class_f1} reports per-class F1 scores for the ten event classes with the largest total duration, while the statistical test uses all 28 classes. A similar pattern is observed for the GeoFusion-Inter-CLAP, with stable F1 scores across Geo-ATBench labels ($F1 = 0.649$) and 10 participant human consensus labels ($F1 = 0.592; W = 187, p > 0.05$). 
Overall, model evaluation remains consistent under Geo-ATBench labels and aggregated human consensus labels on the annotated test set of 579 clips, and the paired Wilcoxon signed-rank tests do not indicate a statistically significant difference between the two label references. This complements the inter-rater reliability analysis and supports Geo-ATBench as a human-aligned benchmark for clip-level evaluation.

\begin{table}[t]
\centering
\footnotesize
\setlength{\tabcolsep}{2pt}
\renewcommand{\arraystretch}{1} 
\setlength{\abovecaptionskip}{0.1cm}   
	\setlength{\belowcaptionskip}{-0.3cm}  
\begin{tabular}{lccc @{\hspace{4pt}}|@{\hspace{4pt}} lccc}
\toprule
Event & F1 (Label) & F1 (Human) & Dur. &
Event & F1 (Label) & F1 (Human) & Dur. \\
\midrule
Bird sounds & 0.861 & 0.856 & 8191 &
Falling water & 0.872 & 0.786 & 2922 \\
Speech & 0.827 & 0.869 & 5133 &
Flowing water & 0.774 & 0.667 & 2774 \\
Plane & 0.779 & 0.514 & 3092 &
Waves & 0.756 & 0.677 & 2754 \\
Crickets & 0.883 & 0.836 & 3091 &
Footsteps & 0.629 & 0.607 & 2225 \\
Car & 0.491 & 0.404 & 3068 &
Musical instru. & 0.651 & 0.667 & 1593 \\
\bottomrule
\end{tabular}
\caption{Top-10 event classes (total = 34{,}843~s $\approx$ 9.68~h) by descending total duration in Geo-ATBench. F1 (Label) and F1 (Human) are F1 scores of GeoFusion-Inter-CLAP predictions evaluated against Geo-ATBench labels and aggregated human consensus labels, respectively. Dur. denotes total event duration (seconds). Musical instru. denotes Musical instrument, and Falling water corresponds to Falling water/Rain.}
\label{tab:class_f1}
\end{table}

\vspace{-2mm}
\section{Conclusion}
\label{section_Conclusion}

\vspace{-2mm}
Environmental sound events do not exist in isolation: they are physical phenomena generated and perceived within specific geographic environments. Nevertheless, computational auditory scene analysis (CASA) often treats multi-label audio tagging as an audio-only inference problem. Under polyphonic overlap, audio information can be insufficient to separate labels with similar acoustic patterns, and disambiguating cues may lie outside the waveform.

In response to this gap, we introduce the Geospatial Audio Tagging (Geo-AT) task, which frames multi-label audio tagging conditioned on paired audio and geospatial semantic context (GSC). Geo-AT focuses on POI-derived, location-tied semantics as complementary priors that are not encoded in the audio signal. This task-level formulation provides a principled foundation for integrating spatial semantics into machine listening, extending the scope of CASA beyond independent signal analysis.

Geo-ATBench is released to support reproducible Geo-AT evaluation. Geo-ATBench contains 3,854 clips (10.71 hours) with 28 event classes of real-world polyphonic audio, and each clip is paired with a POI-derived GSC representation constructed from OSM annotations over 11 semantic context categories. By explicitly encoding the semantic characteristics of recording environments, Geo-ATBench addresses an important resource gap in the field. It enables controlled studies on how spatial context interacts with acoustic representations and offers a shared benchmark for evaluating geospatially grounded sound classification. GeoFusion-AT is introduced to report reference results on Geo-ATBench using feature-level, representation-level, and decision-level fusion with three AudioSet-pretrained backbones, PANNs, AST, and CLAP. Across backbones and fusion points, incorporating GSC is associated with improved mAP on the 28-class Geo-AT task in most configurations. 
 
A crowdsourced listening study with 10 participants further supports Geo-ATBench as a human-aligned reference on the annotated test set of 579 clips. Together, the Geo-AT task, the Geo-ATBench dataset, and the GeoFusion-AT framework provide a concrete basis for studying how location-tied semantics can complement acoustic representations in CASA.

\vspace{-4.7mm}
\bibliographystyle{elsarticle-num}
\bibliography{cas-refs}
\end{document}